\documentclass[aps,prc,preprint,showpacs,showkeys,floatfix]{revtex4}
\usepackage{epsfig}
\usepackage{graphicx}
\usepackage{dcolumn}
\usepackage{bm}

\begin{document}
\title{Phase transitions and critical points in the rare-earth region}
\author{J.E.~Garc\'{\i}a-Ramos}\email{jegramos@nucle.us.es} 
\affiliation
{Departamento de F\'{\i}sica Aplicada, 
Facultad de Ciencias Experimentales, Universidad de Huelva, 
21071 Huelva, Spain} 

\author{J.M.~Arias} 
\affiliation
{Departamento de F\'{\i}sica At\'omica, Molecular
y Nuclear, Universidad de Sevilla, Apartado 1065, 41080 Sevilla, Spain}

\author{J.~Barea} 
\affiliation
{Departamento de F\'{\i}sica At\'omica, Molecular
y Nuclear, Universidad de Sevilla, Apartado 1065, 41080 Sevilla, Spain}
\affiliation
{Centro de Ciencias F\'{\i}sicas, Universidad Nacional 
Aut\'onoma de M\'exico, Apartado Postal 139-B, 62251 Cuernavaca, 
Morelos, M\'exico}

\author{A.~Frank} 
\affiliation
{Instituto de Ciencias Nucleares, 
Universidad Nacional Aut\'onoma de M\'exico, Apartado Postal 70-543, 
04510 M\'exico, DF, M\'exico} 
\affiliation
{Centro de Ciencias F\'{\i}sicas, Universidad Nacional 
Aut\'onoma de M\'exico, Apartado Postal 139-B, 62251 Cuernavaca, 
Morelos, M\'exico}

\date{\today}

\begin{abstract}
A systematic study of isotope chains in the rare--earth region 
is presented. For the chains $^{144-154}_{\phantom{4-154}60}$Nd,
$^{146-160}_{\phantom{6-160}62}$Sm,  
$^{148-162}_{\phantom{8-162}64}$Gd, and 
$^{150-166}_{\phantom{0-166}66}$Dy, energy levels, E2 transition
rates, and two--neutron separation energies are described by
using the most general (up to two--body terms) IBM Hamiltonian. 
For each isotope chain a general fit is performed in such a way that all
parameters but one are kept fixed to describe the whole chain. 
In this region nuclei evolve from spherical to deformed shapes and 
a method based on catastrophe theory,  in combination with a
coherent state analysis to generate the IBM energy surfaces, 
is used to identify  critical phase transition points. 
\end{abstract}

\pacs{21.60.-n, 21.60.Fw}

\keywords{quantum phase transitions, catastrophe theory, 
interacting boson model}

\maketitle

\vspace{1cm}



\newpage

\section{Introduction}
\label{sec-intro}
Recently, a renewed interest in the study of quantum phase transitions
in  atomic nuclei  has emerged  \cite{Rowe98,Iach98,Cast99,Joli99}.  A
new  class of  symmetries that  applies  to systems  localized at  the
critical  points  has been  proposed.  In  particular, the  ``critical
symmetry''  $E(5)$  \cite{Iach00}   has  been  suggested  to  describe
critical   points  in   the   phase  transition   from  spherical   to
$\gamma$-unstable  shapes while  $X(5)$ \cite{Iach01}  is  designed to
describe systems  lying at the  critical point in the  transition from
spherical to  axially deformed systems. These are  based originally on
particular solutions of the Bohr-Mottelson differential equations, but
are  usually applied  in the  context of  the Interacting  Boson Model
(IBM) \cite{Iach87},  since the latter provides a  simple but detailed
framework in  which first  and second order  phase transitions  can be
studied. In the  IBM language, the symmetry $E(5)$  corresponds to the
critical point  between the $U(5)$  and $O(6)$ symmetry  limits, while
the  $X(5)$  symmetry  should  describe the  phase  transition  region
between the $U(5)$ and  the $SU(3)$ dynamical symmetries, although the
connection  is not  a rigorous  one.  Very  recently the  $O(6)$ limit
itself  has also  been  proposed  to correspond  to  a critical  point
\cite{Joli01}.

Usually, the IBM  analyses of phase transitions have  been carried out
using schematic Hamiltonians in which the transition from one phase to
the other is  governed by a single parameter. It  is thus necessary to
see how much these predictions vary when a more general Hamiltonian is
used. The global approach was first used by Casta\~nos {\it et al} for
the study of series of isotopes \cite{Cast82,Fran89,Gome95}. An
alternative  procedure is  provided by  the  use of  the consistent  Q
formalism (CQF) \cite{Warn82}. In  this case, although the Hamiltonian
is   simpler  than  the   general  one,   the  main   ingredients  are
included. Within  this scheme  a whole isotope  chain is  described in
terms of few  parameters that change smoothly from  one isotope to the
next.  Because  of the  possible  non-uniqueness  of  such nucleus  by
nucleus fits  and the restricted  parameter space, it is  important to
study  under what  circumstances  the prediction  of  the location  of
critical  points in a  phase transition  is robust.  In this  paper we
follow  Refs.~\cite{Cast82,Fran89,Gome95,Lope96,Lope98} and use
a more general one-- and two--body IBM Hamiltonian to obtain the model
parameters from a fit to energy  levels of chains of isotopes. In this
way a set  of fixed parameters, with the exception  of one that varies
from isotope  to isotope, is obtained  for each isotope  chain and the
transition phase can  be studied in the general  model space.  The fit
to a large data set in many nuclei diminishes the uncertainties in the
parameter determination.  A possible problem arising from working with
such a general Hamiltonian,  however, is the difficulty in determining
the  position of  the  critical points.  Fortunately,  the methods  of
catastrophe theory \cite{Gilm81} allow the definition of the essential
parameters needed  to classify the  shape and stability of  the energy
surface \cite{Lope96,Lope98}.

In this paper we analyze diverse  spectroscopic properties of
several isotope chains in the rare-earth
region,  in which shape transition from spherical to deformed shapes is
observed. We combine this study with a coherent-state analysis and
with 
 catastrophe theory in order to localize the critical points and test
the $X(5)$ predictions. Since  the introduction of
the  $E(5)$ and $X(5)$ symmetries,   only a small number of candidates 
\cite{Zamf99,Cast00,Klug00,Cast01,Fran01,Zamf02,Bizz02,Kruc02}  
have been proposed as possible realizations of such  critical point
symmetries. In this paper we  
show that the critical points can be clearly identified by means of  a general
theoretical approach \cite{Lope96,Lope98}.

The paper is structured as follows. 
In section \ref{sec-ibm} we present the IBM Hamiltonian used. In section
\ref{sec-fits} the results of the fits made for the different isotope
chains  are presented. Comparisons of the theoretical results 
with the experimental data for
excitation energies, E2 transition rates and two-neutron
separation  energies are shown. 
In section \ref{sec-surf} the intrinsic state formalism is
used to generate the energy surfaces produced by the parameters
obtained in the preceding section. In addition, the location of the critical
point in the shape transition for each isotope chain is identified by
using  catastrophe theory. Also in this section, the alternative
description provided by the CQF for the rare-earth region is briefly
discussed. Finally, section \ref{sec-conclu} is
devoted to summarize and to present  our  conclusions.

\section{IBM description}
\label{sec-ibm}
In this work we use the interacting boson model (IBM) to study in a
systematic way the properties of the low-lying nuclear collective
states in several even--even isotope chains in the rare-earth region.
The building blocks of the model are bosons with
angular momentum $L=0$ ($s$ bosons) and $L=2$ ($d$
bosons). The dynamical  algebra of the model is $U(6)$. Therefore, every
dynamical operator, such as the Hamiltonian or the transition
operators, can be written in terms of the generators of the 
latter algebra. Usually some restrictions are imposed on these
operators, {\it e.g.}~the
Hamiltonian should be number conserving and rotational invariant, and 
in most cases it only includes up to two-body terms. 

The most general (including up to two--body terms) IBM Hamiltonian,
using the multipolar form, can be written as
\begin{eqnarray}
\label{ham1}
\hat H&=&\tilde{\cal A} \hat N+ \tilde{\cal B} {\hat N (\hat N-1)\over 2}+
\varepsilon_d \hat n_d +
\kappa_0 \hat P^\dag \hat P\nonumber\\
&+&\kappa_1 \hat L\cdot \hat L+
\kappa_2 \hat Q \cdot \hat Q+
\kappa_3 \hat T_3\cdot\hat T_3 +\kappa_4 \hat T_4\cdot\hat T_4
\end{eqnarray}  
where $\hat N$, and $\hat n_d$ are the total boson number operator,
and the $d$ boson number operator, respectively and 
\begin{eqnarray}
\label{P}
\hat P^\dag&=&\frac{1}{2} ~ (d^\dag \cdot d^\dag - s^\dag \cdot s^\dag), \\
\label{L}
\hat L&=&\sqrt{10}(d^\dag\times\tilde{d})^{(1)},\\
\label{Q}
\hat Q&=& (s^{\dagger}\times\tilde d
+d^\dagger\times\tilde s)^{(2)}-
{\sqrt{7}\over 2}(d^\dagger\times\tilde d)^{(2)},\\
\label{t3}
\hat T_3&=&(d^\dag\times\tilde{d})^{(3)},\\
\label{t4}
\hat T_4&=&(d^\dag\times\tilde{d})^{(4)}.
\end{eqnarray} 
The symbol $\cdot$ stands for the scalar product,    defined as 
$\hat T_L\cdot \hat T_L=\sum_M (-1)^M \hat T_{LM}\hat T_{L-M}$ where 
$\hat T_{LM}$ corresponds to the $M$ component of the operator 
$\hat T_{L}$. The operator 
$\tilde\gamma_{\ell m}=(-1)^{m}\gamma_{\ell -m}$ (where $\gamma$
refers to $s$ and $d$ bosons) is introduced to ensure the correct
tensorial character under spatial rotations.

The first two terms in the
Hamiltonian do not affect the spectra but only the binding
energy. Therefore they can be removed from the Hamiltonian if only the
excitation spectrum of the system is of interest. However, a complete
description of both excitation and binding energies requires the use
of the full Hamiltonian (\ref{ham1}).

The electromagnetic transitions can also be analyzed in the framework of
the IBM. In particular, in this work we will focus 
on $E2$ transitions. The
most general $E2$ transition operator including up to one body terms 
can be written as
\begin{equation}
\label{te2}
\hat T^{E2}_M=e_{eff}\left[(s^\dag \times \tilde{d}+d^\dag\times\tilde{s})^{(2)}_M+
\chi(d^\dag\times\tilde{d})^{(2)}_M\right],
\end{equation}
where $e_{eff}$ is the boson effective charge and $\chi$ is a structure 
parameter.

Two-neutron separation energies (S$_{2n}$) are also studied in the present work. This observable is defined as the difference in binding energy
between an even-even isotope and the preceding even-even one,
\begin{equation}
\label{s2n}
S_{2n}=BE(N)-BE(N-1), 
\end{equation} 
where $N$ corresponds to the total number of valence bosons. Note that if
only the first two terms in (\ref{ham1}) are considered and
$\tilde{\cal A}$ and $\tilde{\cal B}$ are assumed to be constant along
the isotope chain, S$_{2n}$ would be given by 
\begin{equation}
\label{s2n-lin}
S_{2n}=-(\tilde{\cal A}-{1\over 2}\tilde{\cal B})-\tilde{\cal B}N 
      = {\cal A}+{\cal B}N .
\end{equation} 
For a detailed study of this property we refer to Ref. \cite{Foss02a}. 

\section{Fits}
\label{sec-fits}
In this section we analyze several isotope chains belonging
to the rare-earth region using the most general IBM Hamiltonian,
Eq.~(\ref{ham1}),  and  $E2$ transition operator, Eq.~(\ref{te2}). As
an {\it ansazt}  for each chain of isotopes we will assume a single 
Hamiltonian, and a single $E2$ transition operator. All   parameters
in these operators are kept fixed for a given  isotope chain,
except for the single particle energy which is allowed to
vary slightly from isotope to isotope. 
The way of fixing the best set of parameters in the Hamiltonian is to
carry out a least-square fit procedure of  the excitation energies
of selected states ($2^+_1$, $4^+_1$, 
$6^+_1$, $8^+_1$, $0^+_2$, $2^+_3$, $4^+_3$, $2^+_2$, $3^+_1$, and  
$4^+_2$) and the two neutron separation energies of all isotopes in
each isotopic  chain. 
Once the parameters in the Hamiltonian are obtained, the $B(E2)$ 
transition probabilities $2^+_1\rightarrow 0^+_1$, $4^+_1\rightarrow
2^+_1$, $2^+_2\rightarrow 0^+_1$, $2^+_3\rightarrow 0^+_1$,
$0^+_2\rightarrow 2^+_1$, and $0^+_3\rightarrow 2^+_1$ of the set of  isotopes
are used to fix  $e_{eff}$ and $\chi$ by carrying out a least-square fit. 
The experimental data for excitation and binding
energies and B(E2)'s have been taken from Refs.~
\cite{Sonz01,Peke97,Bhat00,Derm95,Artn96,Reic98,Helm92,Helm96,Reich96,Helm99,Balr01,Shur92,Audi95}. 
Finally, it is worth noting that in Ref.~\cite{Foss02a} the
Hamiltonian parameters were fixed just using the data for excitation
energies and then ${\cal A}$ and ${\cal B}$ were adjusted to reproduce
the experimental values of S$_{2n}$. In this paper, since we are 
particularly interested in accurately describing the spectroscopic 
data associated to shape transitions, both, excitation
and binding energies, are treated on an  equal footing describing the 
shape transition,  to determine the set of 
Hamiltonian parameters in Eq. (\ref{ham1}).

Tables \ref{tab-ham0} and  \ref{tab-ham}  summarize  the parameters 
obtained for  
the Hamiltonian and $E2$ transition operator for  each isotope chain.  

\begin{figure}[tbt]
\begin{center}
\mbox{\epsfig{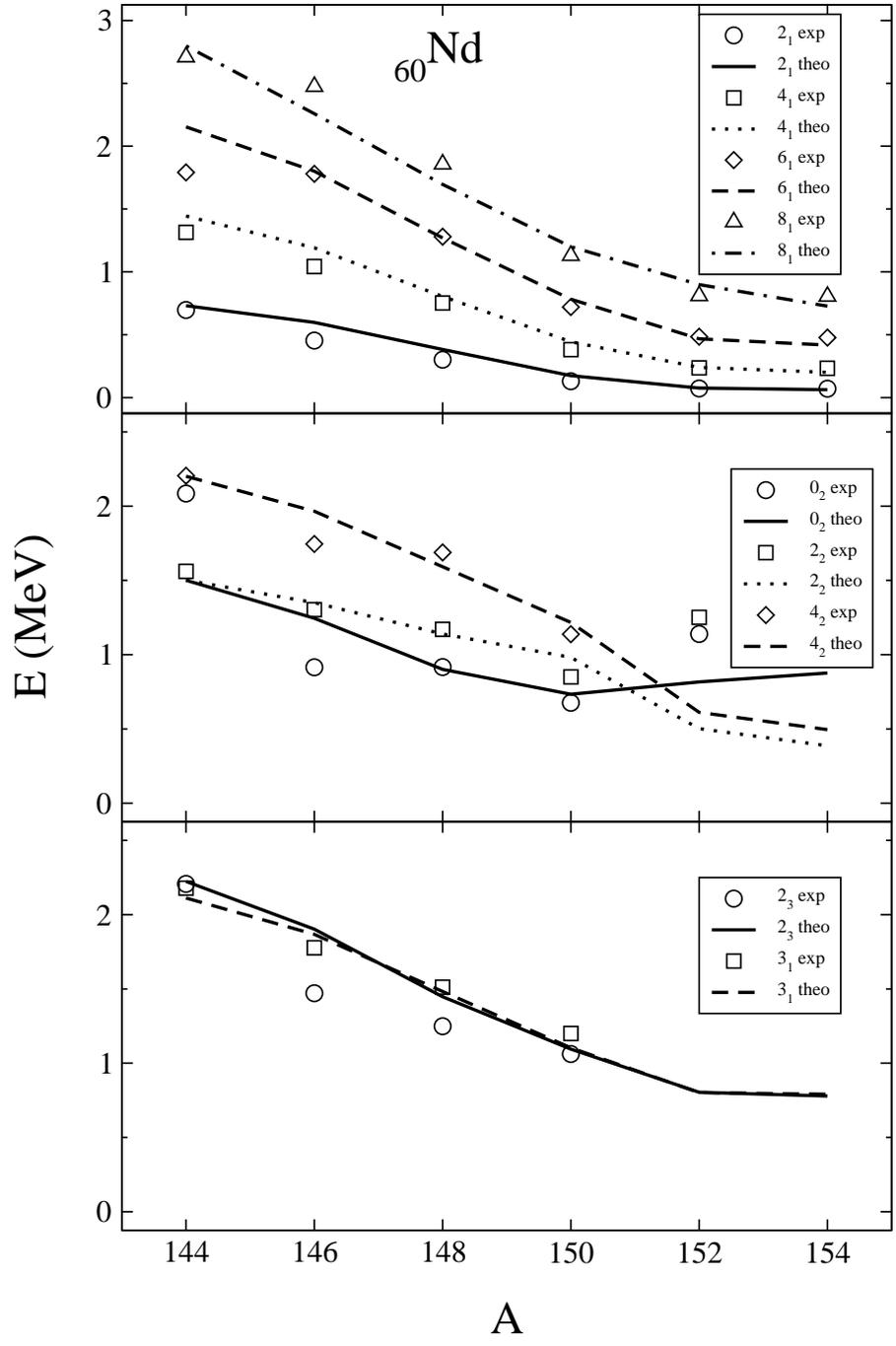}}
\end{center}
\caption{Excitation energies of Nd isotopes.}
\label{fig-ener-nd}
\end{figure}

\begin{figure}[tbt]
\begin{center}
\mbox{\epsfig{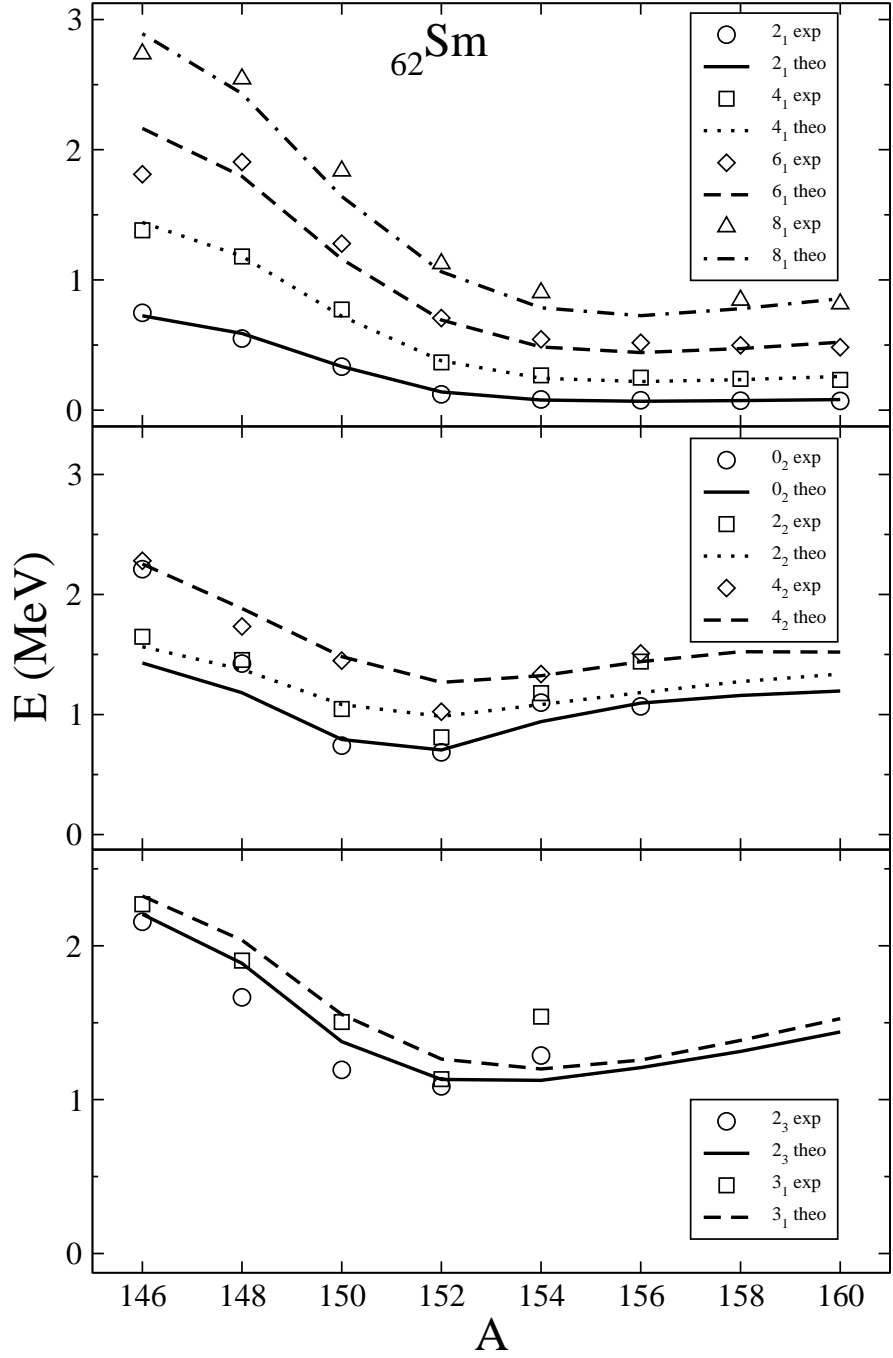}}
\end{center}
\caption{Excitation energies of Sm isotopes.}
\label{fig-ener-sm}
\end{figure}

\begin{figure}[tbt]
\begin{center}
\mbox{\epsfig{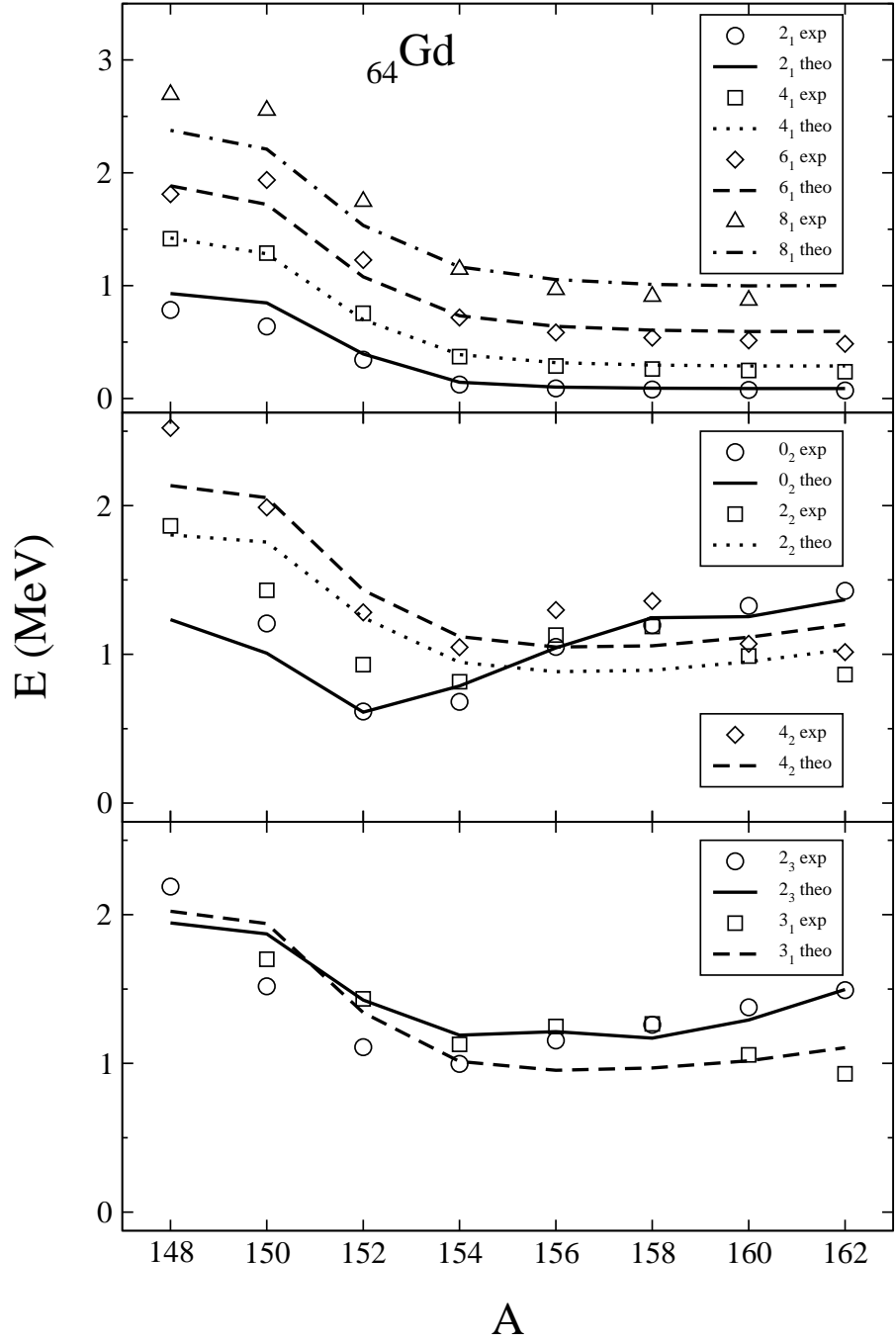}}
\end{center}
\caption{Excitation energies of Gd isotopes.}
\label{fig-ener-gd}
\end{figure}

\begin{figure}[tbt]
\begin{center}
\mbox{\epsfig{file=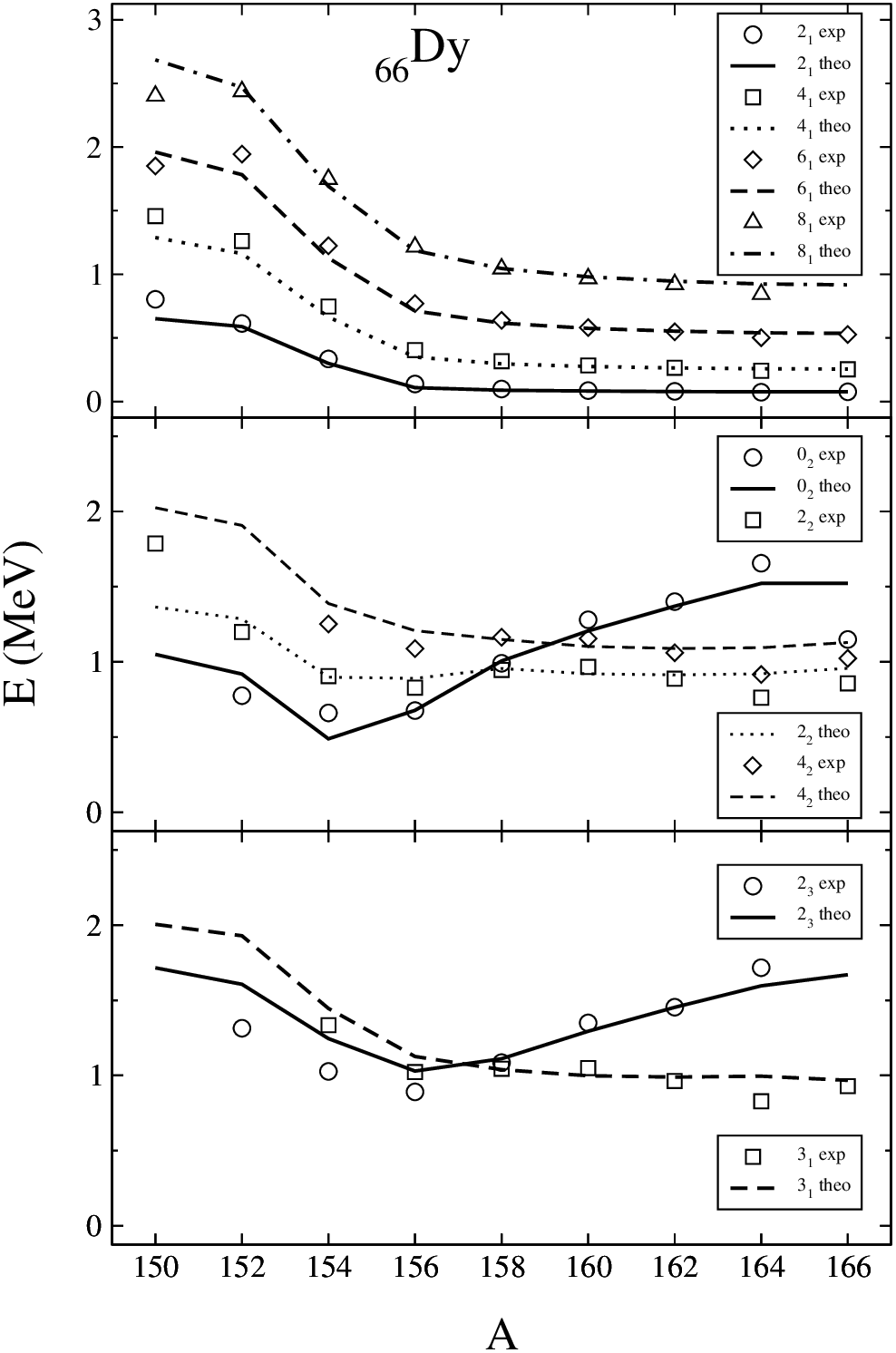,height=18.0cm,angle=0}}
\end{center}
\caption{Excitation energies of Dy isotopes.}
\label{fig-ener-dy}
\end{figure}

\begin{figure}[tbt]
\begin{center}
\mbox{\epsfig{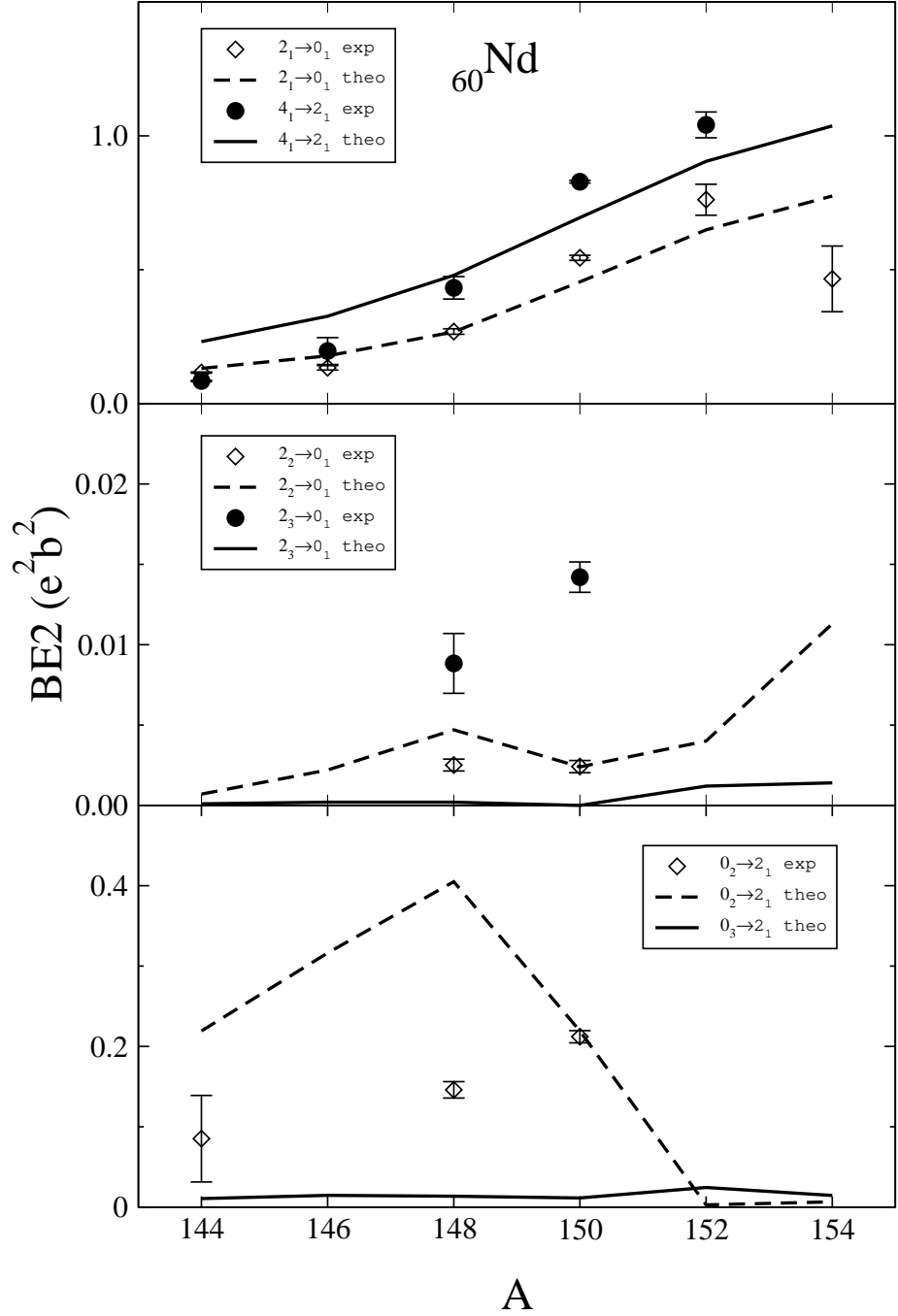}}
\end{center}
\caption{$B(E2)$ transition rates for Nd isotopes.}
\label{fig-be2-nd}
\end{figure}

\begin{figure}[tbt]
\begin{center}
\mbox{\epsfig{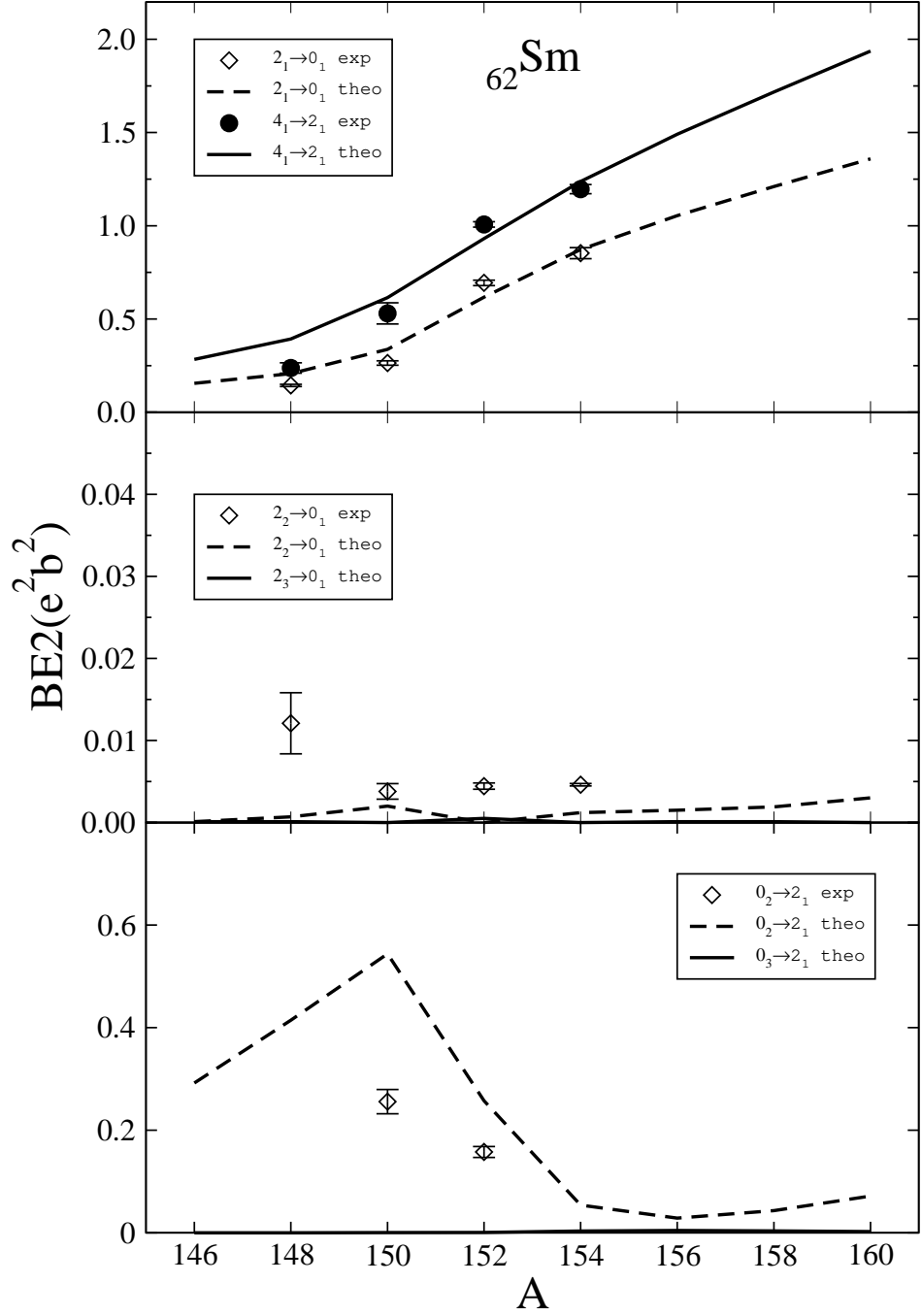}}
\end{center}
\caption{$B(E2)$ transition rates for Sm isotopes.}
\label{fig-be2-sm}
\end{figure}

\begin{figure}[tbt]
\begin{center}
\mbox{\epsfig{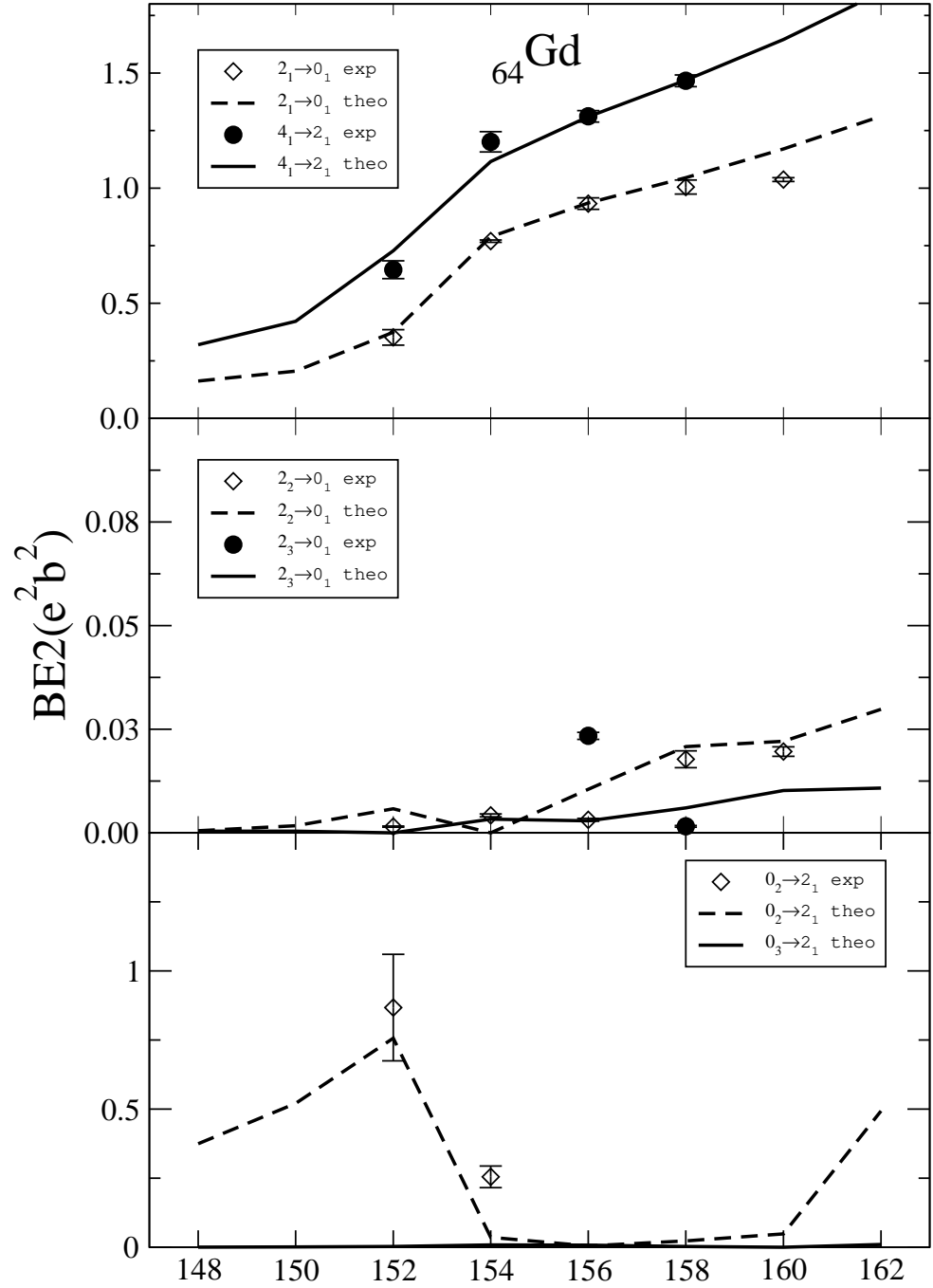}}
\end{center}
\caption{$B(E2)$ transition rates for Gd isotopes.}
\label{fig-be2-gd}
\end{figure}

\begin{figure}[tbt]
\begin{center}
\mbox{\epsfig{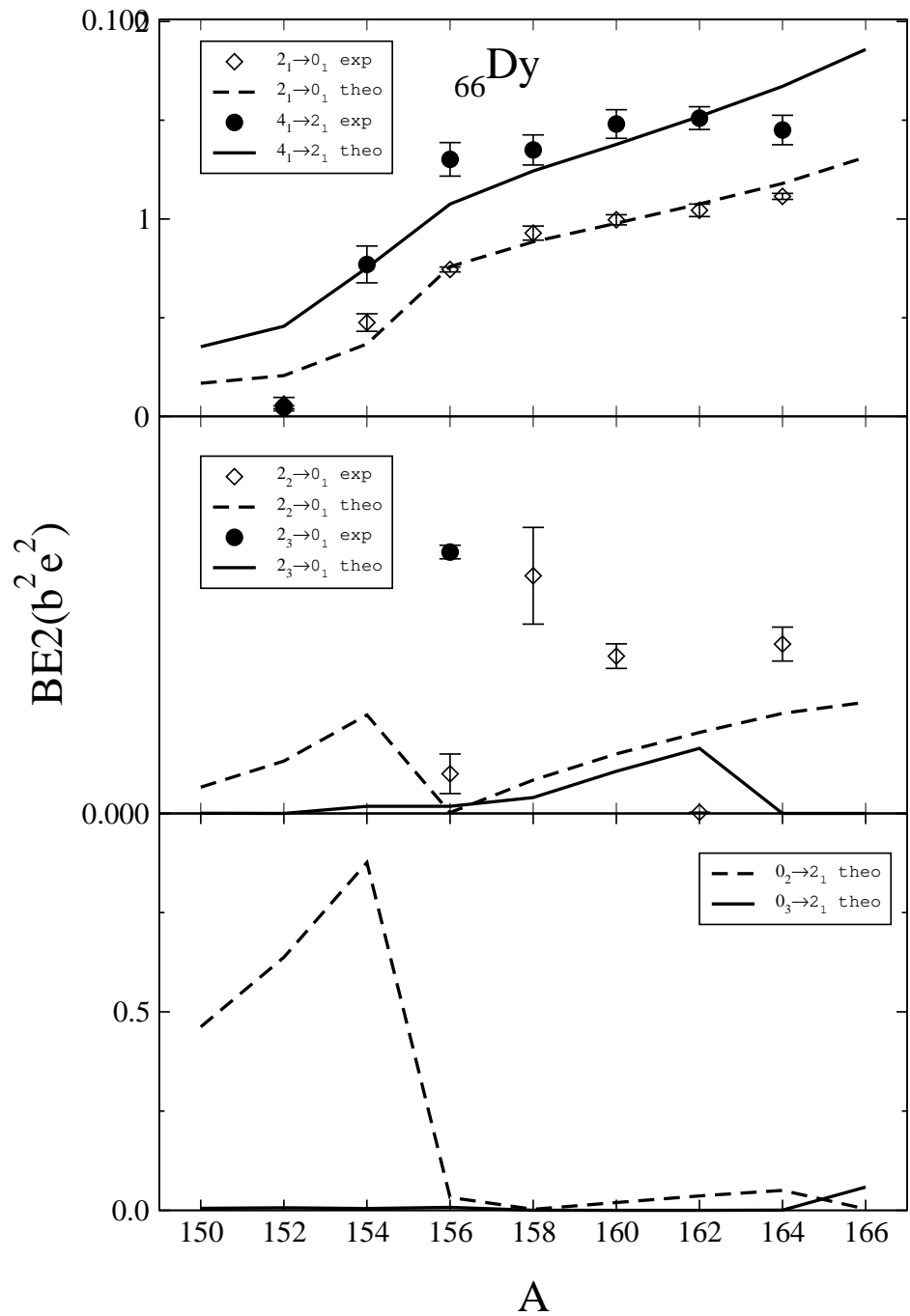}}
\end{center}
\caption{$B(E2)$ transition rates for Dy isotopes.}
\label{fig-be2-dy}
\end{figure}

\begin{figure}[tbt]
\begin{center}
\mbox{\epsfig{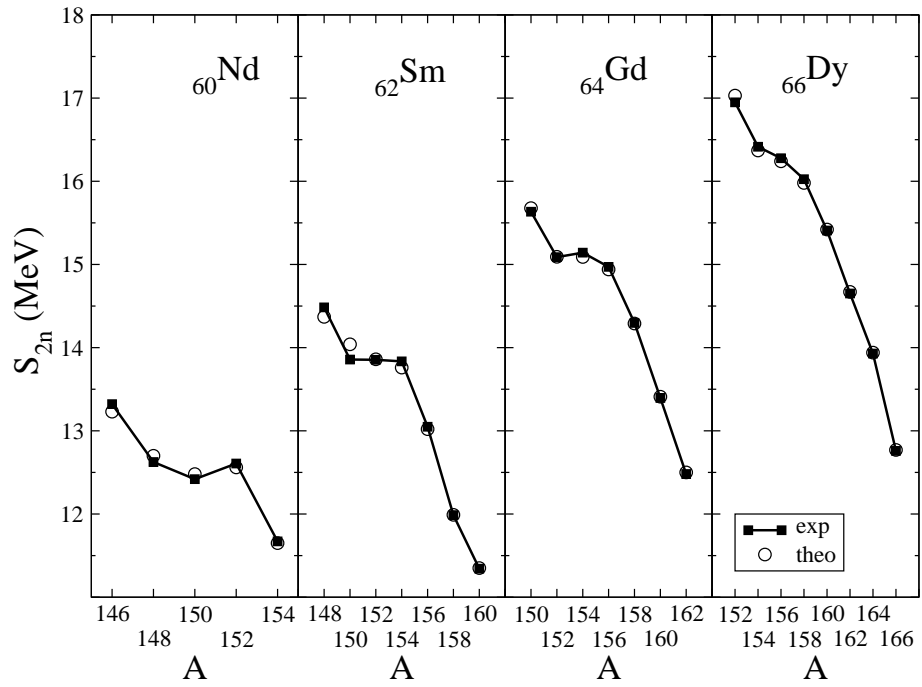}}
\end{center}
\caption{S$_{2n}$ values for Nd, Sm, Gd, and Dy isotopes.}
\label{fig-s2n}
\end{figure}

In figures \ref{fig-ener-nd}, \ref{fig-ener-sm}, \ref{fig-ener-gd},
and \ref{fig-ener-dy} the systematics of experimental and calculated
energies for the states included in the least-square procedure are
presented in order to show the goodness of the fitting procedure.  
In figures \ref{fig-be2-nd}, \ref{fig-be2-sm}, \ref{fig-be2-gd},
and \ref{fig-be2-dy} the systematics of the experimental and calculated
$B(E2)$ values are compared. 
Finally, in figure \ref{fig-s2n} the experimental and
calculated S$_{2n}$ values are shown. This is a fundamental magnitude
for identifying a phase transition since it is directly related to the
derivative of the energy surface. First order phase transitions 
are related with the appearance of a kink 
in the S$_{2n}$ values. As shown in
Fig. \ref{fig-s2n}, the calculation matches  the experimentally observed
behavior. 

The analysis of the preceding figures for different observables
and for several isotope chains shows that the
present procedure is appropriate for systematic studies 
and confirms that it provides a simple framework to describe 
long chains of isotopes and detect possible phase transitions.

\begin{figure}[tbt]
\begin{center}
\mbox{\epsfig{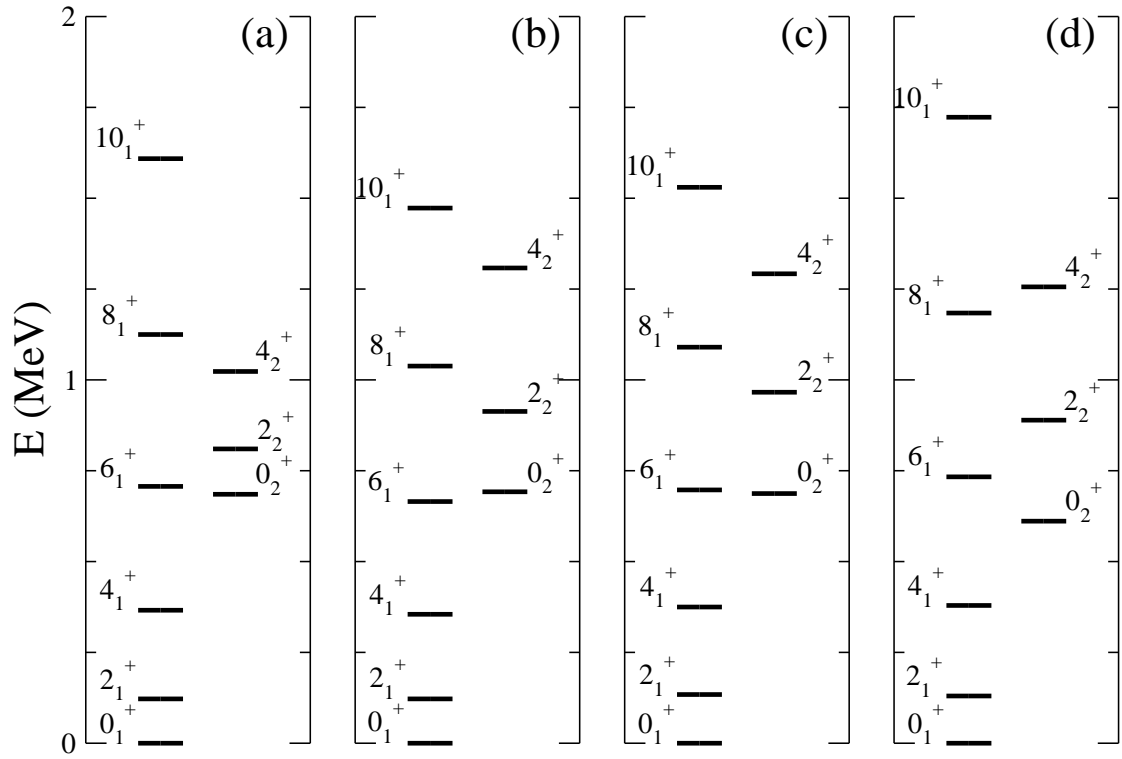}}
\end{center}
\caption{Spectrum of $^{152}$Sm: (a) experimental, (b) X(5) symmetry,
  (c) this work, and (d) using CQF \cite{Iach98}.}
\label{fig-spect-152sm}
\end{figure}

An alternative approach to describe long chains of rare-earth nuclei is
to use the CQF. The CQF Hamiltonian is
\begin{equation}
\hat{H}= \epsilon \hat n_d + \kappa \hat Q^\prime \cdot \hat Q^\prime ~,
\end{equation}
with 
\begin{equation}
\hat Q^\prime = (s^{\dagger}\times\tilde d
+d^\dagger\times\tilde s)^{(2)} + \chi (d^\dagger\times\tilde d)^{(2)}.
\end{equation}
For each nucleus the parameters $\epsilon$, $\kappa$ and $\chi$ are
determined in order to fit the excitation energies and B(E2)'s.
In particular in Ref.~\cite{Chou97} the
parameters of the Hamiltonian are calculated within the CQF framework
with the {\it ansazt} that the strength of the quadrupole term of the
Hamiltonian remains  constant along a wide region of the mass table. 
 As in  the present paper they compare
experimental data and theoretical values for excitation
energies and $B(E2)$ transition rates.    
Both methods provide a consistent description of the rare-earth
region with a similar number of parameters as can be observed in
Fig.~\ref{fig-spect-152sm} and in table \ref{tab-be2-152sm} where the
case of $^{152}$Sm is analyzed. Note that in the present work the results
come from a global analysis, therefore the $B(E2)$ transition rates
are not normalized to the transition $B(E2:2_1^+\rightarrow
0_1^+)$ in a particular isotope. If in table \ref{tab-be2-152sm} the
results are normalized so as to reproduce the observed value for
$B(E2:2_1^+\rightarrow 0_1^+)$ in $^{152}$Sm the results of this work and
CQF are basically the same.

\section{Energy surfaces and phase transitions}
\label{sec-surf}
The study of phase transitions in the IBM requires the use of the so 
called intrinsic-state formalism \cite{Gino80,Diep80,Diep80b}, 
although other approaches can be used
\cite{Cast99,Cejn00}. This formalism is very useful to discuss phase
transitions in finite systems because it provides a   description of 
the behavior of a macroscopic system up to $1/N$ effects. 
To define the intrinsic, or coherent, state it is
assumed that the dynamical behavior of the system can be described in
terms of independent bosons (``dressed bosons'') moving in an average
field \cite{Duke84}. The ground state of the system is a condensate,
$|c\rangle$, of bosons occupying the lowest--energy phonon state,
$\Gamma^\dag_c$,
\begin{equation}
\label{GS}
| c \rangle = {1 \over \sqrt{N!}} (\Gamma^\dagger_c)^N | 0 \rangle,
\end{equation}
where
\begin{equation}
\label{bc}
\Gamma^\dagger_c = {1 \over \sqrt{1+\beta^2}} \left (s^\dagger + \beta
\cos     \gamma          \,d^\dagger_0          +{1\over\sqrt{2}}\beta
\sin\gamma\,(d^\dagger_2+d^\dagger_{-2}) \right)
\end{equation}
and
$\beta$ and $\gamma$ are variational parameters related with the shape
variables in the geometrical collective model. The expectation value
of the Hamiltonian in the intrinsic  state (\ref{GS}) provides  
the energy surface of the system, 
$E(N,\beta,\gamma)=\langle c|\hat H| c \rangle$.  
The energy surface in terms of the
parameters of the Hamiltonian (\ref{ham1}) and the shape variables can be 
readily obtained \cite{Isac81},
 \begin{eqnarray}
\label{Ener1}
\langle c|\hat H | c \rangle&=&
{\displaystyle {N\beta^2\over{(1+\beta^2 )}}}
\Bigl(\varepsilon_d+
6 \,\kappa_1 
-{9\over 4}\,\kappa_2+{7\over 5}\,\kappa_3+{9\over 5}
\,\kappa_4\Bigr)\nonumber\\
&+&{\displaystyle {\frac{N(N-1)}{{{(1+\beta^2)}^2}}}}
\Big[\frac{\kappa_0}{4}+\beta^2(- \frac{\kappa_0}{2}+4\,\kappa_2)
+2\,{\sqrt{2}}\,\beta^3\,\kappa_2\,\cos(3\,\gamma)
\nonumber\\
\qquad\qquad\qquad
&&+\beta^4(\frac{\kappa_0}{4}
+\frac{\kappa_2}{2}+\frac{18}{35}\,\kappa_4)\Big],
\end{eqnarray}
where the terms which do not depend on $\beta$ and/or $\gamma$
(corresponding to $\tilde{\cal A}$ and $\tilde{\cal B}$ in 
Eq.~(\ref{ham1})) have not been included. 

The equilibrium values of the variational parameters $\beta$ and
$\gamma$ are obtained by minimization of the ground state energy
$\langle c|\hat H| c \rangle$. 
As mentioned above these parameters are related to
the  parameters of the Geometrical Collective Model and
provide an image of the nuclear shape for a given IBM Hamiltonian.
A spherical nucleus has a minimum in the energy surface at
$\beta=0$, while for a deformed one the energy surface has a
minimum at a finite value of $\beta$ and $\gamma=0$ (prolate
nucleus) or $\gamma=\pi/3$ (oblate nucleus). Finally, a
$\gamma$-unstable nucleus corresponds to the case in which the energy
surface has a minimum at a particular value of $\beta$ 
and is independent of the value of $\gamma$.
The equilibrium values of $\beta$ and $\gamma$ are the order parameters
to study the phase transition of the system, although in the case
under consideration (IBM-1) only $\beta$ has to be taken into account,
since the minima in $\gamma$ are well defined.

In Fig. \ref{fig-ener} the energy
surfaces for the isotopes of the different isotope chains studied in
this paper are plotted as a function of $\beta$. The figure on the
right is a zoom of the region close to $\beta=0$. 

The classification of phase transitions that  we follow in this paper 
and that  is followed traditionally in the  IBM is the Ehrenfest
classification \cite{Stan71}. In this context, the origin of a phase
transition resides in the way the energy surface (their minima
positions) is changing as a function of the control parameter that, 
in this work, is a combination of parameters of the Hamiltonian (see
Eq.~(\ref{r-line})). 
First order phase transitions appear when 
there exists a discontinuity in the first derivative of the
energy with respect to the control parameter. This discontinuity 
appears when two
degenerate  minima exist in the energy surface for two values of the
order parameter $\beta$. Second order phase
transitions appear when the second derivative of the energy with
respect to the control parameter displays  a discontinuity. This
happens when the energy surface presents  a single minimum for 
$\beta=0$ and the surface satisfies the condition  
${\left(d^2 E\over d\beta^2\right)_{\beta=0}}=0$.
 
With the introduction of the  $E(5)$ and
$X(5)$ symmetries to describe phase 
transitional  behavior,
 diverse  attempts to identify nuclei that could be located at
the critical points have been made. The theoretical approaches have
been mainly performed with restricted IBM Hamiltonians. 
In particular, within
the CQF, or other restricted Hamiltonians,
the location of the critical point is obtained by 
imposing ${d^2 E\over d\beta^2}=0$ at  $\beta=0$, 
where $E$ is the energy surface \cite{Iach98}.  
This condition leads to a flat surface in a   region of small values 
of $\beta$,
with a single minimum in the limit $\chi=0$ and two
almost degenerate minima (one of them in $\beta=0$) in the other cases. 
In the CQF approximation it can be said that 
${\left(d^2 E\over d\beta^2\right)_{\beta=0}}=0$ corresponds
approximately to a ``very flat energy surface'' as happens for 
the $E(5)$ and $X(5)$ critical point  models.  Following this
approach both  $^{150}$Nd and $^{152}$Sm have been found to be  close to
critical. 
However, when studying a transitional region in which the lighter
nuclei are spherical and the heavier are well deformed, 
the a priori  restriction of the parameter space could play a crucial 
role in  the identification of a
particular isotope as critical. It is thus  
important to perform a general analysis in order to
check whether the predictions obtained within the CQF for those nuclei 
close to a  critical point
are robust. We present below such an analysis in the region of the
rare-earths. 
We follow closely the approach introduced
in Ref.~\cite{Lope96,Lope98} using  catastrophe theory. In the next
subsection the main ingredients of the theory 
are summarized and the relevant equations are particularized for the
IBM Hamiltonian written in multipolar form, Eq.~(\ref{ham1}).

\subsection{The separatrix plane}
\label{sec-sepa}
For the study of phase transitions in the  IBM within the framework of 
  catastrophe theory we already have the basic ingredients: 
the Hamiltonian of the system,
Eq. (\ref{ham1}), and the intrinsic state,
Eq. (\ref{GS}).  With them, we have generated the corresponding energy
surface, Eq. (\ref{Ener1}), in terms of the Hamiltonian parameters and
the shape variables.
It is our purpose to find the values of the parameters of the
Hamiltonian that correspond to critical points. In principle this
analysis involves the $6$ parameters of the Hamiltonian, but a first 
simplification occurs since the
energy surface only depends on $5$ parameters 
:
\begin{eqnarray}
\label{Ener2}
\langle c|\hat H | c \rangle&=&{N\tilde\varepsilon
\beta^2\over{(1+\beta^2 )}}
+{N(N-1)\over (1+\beta^2)^2}
\Big(a_1\beta^4+a_2\beta^3\cos(3\,\gamma)+a_3\beta^2+{u_0\over 2} \Big),
\end{eqnarray}
where
\begin{eqnarray}
\tilde\varepsilon&=& \varepsilon_d+
6\,\kappa_1 
-{9\over 4}\,\kappa_2+{7\over 5}\,\kappa_3+{9\over 5}
\,\kappa_4\nonumber\\
a_1&=&{1\over 4}\,\kappa_0+{1\over 2}\,\kappa_2+{18\over 35}\,\kappa_4
\nonumber\\
a_2&=&2\,\sqrt{2} \,\kappa_2
\nonumber\\
a_3&=&-{1\over 2}\,\kappa_0+4\,\kappa_2
\nonumber\\
u_0&=&{\kappa_0\over2}.
\label{coeff}
\end{eqnarray}
Fortunately, it is 
possible to reduce  the number of relevant (or essential)  parameters
to just two and  study all phase transitions by using
  catastrophe theory \cite{Gilm81}. We refer  the reader
to Refs. \cite{Lope96,Lope98}
for details of the application of this theory to the IBM case. The
idea is to analyze the energy surface and obtain all equilibrium
configurations, {\it i.e.}~to find all the critical points of
Eq. (\ref{Ener2}). First, the critical point of maximum degeneracy has
to be identified. In our case, it corresponds to $\beta=0$. Next, the
bifurcation and Maxwell sets are constructed \cite{Gilm81,Lope96}. 
Finally, the separatrix
of the IBM is obtained by the union of Maxwell and bifurcation sets.
In general a bifurcation set, corresponding to minima, limits an area where
two minima in the energy surface coexist. A second order phase
transition   develops when these  minima become the same.  
The crossing of a Maxwell set
corresponding to minima
leads to a first order phase transition.

In order to follow this scheme, one has to
identify the catastrophe germ of the IBM, 
which is the first term in the expansion of
the energy surface around the critical point of maximum degeneracy 
that cannot be canceled by an arbitrary selection of parameters. In our
case, one finds that the first derivative in $\beta=0$ is always
$0$ because of the critical character of the point for any value of the
parameters. The second and third derivatives can also be canceled with
an appropriate selection of parameters. However, if 
one imposes the cancellation of the fourth derivative, the energy
becomes a constant for any value of $\beta$. This means that the
catastrophe germ is $\beta^4$ and the number
of essential parameters is equal to two, which can be  defined, following 
reference \cite{Lope96,Lope98}, as
\begin{equation}
\label{r1}
r_1={a_3-u_0+\tilde\varepsilon/(N-1)\over 2a_1+
  \tilde\varepsilon/(N-1)-a_3}, 
\end{equation}  

\begin{equation}
\label{r2}
r_2=-{2a_2\over 2a_1+
  \tilde\varepsilon/(N-1)-a_3}, 
\end{equation}  
where $\tilde\varepsilon$, $a_1$, $a_2$, and $a_3$ are defined in
  (\ref{coeff}). The denominator in  both expressions fixes 
the energy scale,
which  means that when it becomes negative, the energy surfaces are
inverted. The essential parameters 
$r_1$ and $r_2$ can also be written in terms of the
parameters appearing in (\ref{ham1}) as ,
\begin{equation}
\label{r1p}
r_1={\tilde\varepsilon- (N-1) (\kappa_0 - 4 \kappa_2)
\over \tilde\varepsilon+ (N-1) (\kappa_0 - 3 \kappa_2 +
\frac{36}{35}\kappa_4)}, 
\end{equation} 

\begin{equation}
\label{r2p}
r_2= - {4 \sqrt{2}\kappa_2 (N-1)
\over \tilde\varepsilon+ (N-1) (\kappa_0 - 3 \kappa_2 +
\frac{36}{35}\kappa_4)}.
\end{equation} 

A property of the parametrization used in this work is that the
different chains of isotopes are located on a straight line that
crosses the point corresponding to the $U(5)$ limit. The equation of
this line is given by 
\begin{equation} 
\label{r-line}
r_1={2\kappa_0-7\kappa_2+{36\over 35}\kappa_4\over
  4\sqrt{2}\kappa_2}r_2 +1
\end{equation} 

\begin{figure}[tbt]
\begin{center}
\mbox{\epsfig{file=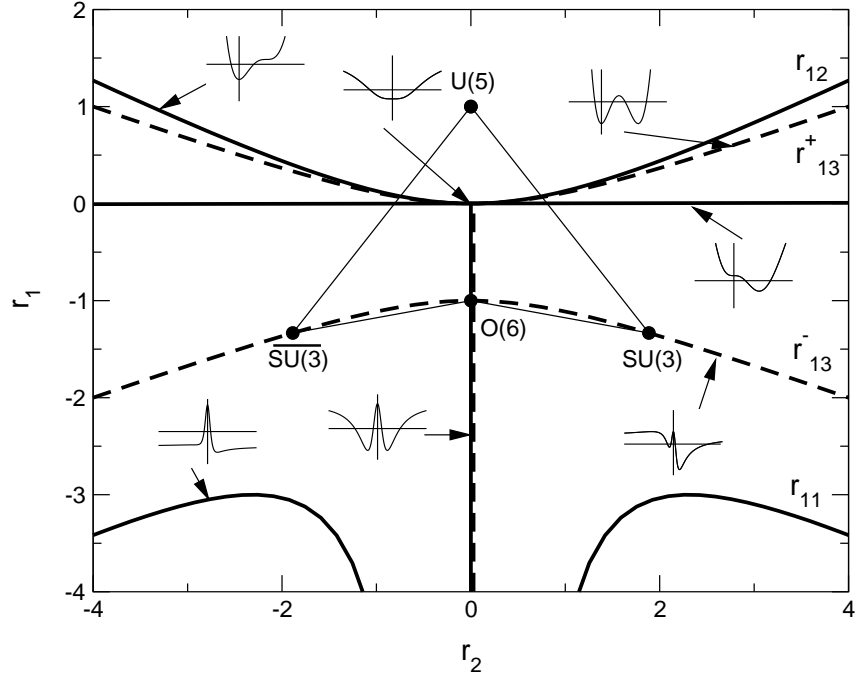,height=9cm,angle=0}}
\end{center}
\caption{Separatrix plane with a positive energy scale.}
\label{fig-cast1}
\end{figure}

\begin{figure}[tbt]
\begin{center}
\mbox{\epsfig{file=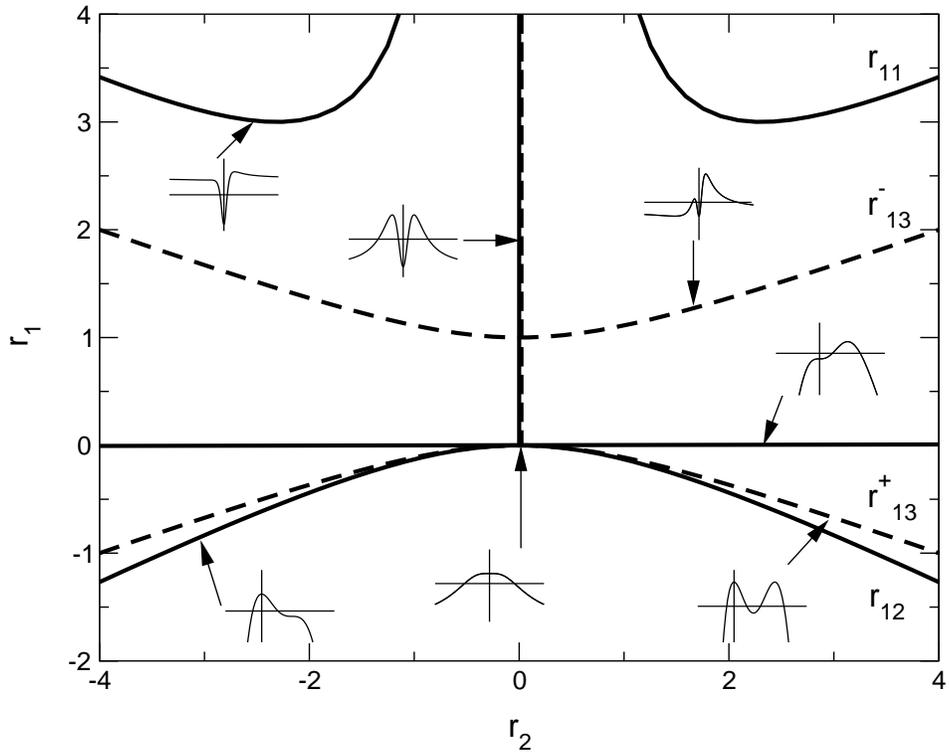,height=10.0cm,angle=0}}
\end{center}
\caption{Separatrix plane with a negative energy scale.}
\label{fig-cast2}
\end{figure}

It should be remarked that the derivation of the essential parameters has
nothing to do with  catastrophe theory. 
The application  of this theory begins once
those parameters are obtained. The basic point is to translate every  set of
Hamiltonian parameters to the plane formed by the essential parameters
$r_1$ and $r_2$. This plane is divided into several sectors by the
bifurcation set, that form the
geometrical place in the parameter space where ${d^2 E\over d
\beta^2}=0$ for a critical value of $\beta$, and the Maxwell sets,
the geometrical place in the space of parameters where two or
more critical points are degenerate \cite{Gilm81}. 
Both sets form the separatrix of the system, in this case of the  
IBM. 
In Ref.~\cite{Lope96,Lope98} the IBM bifurcation  
($r_2$ axis, $r_2=0$  and
$r_1<0$ semi-axis, $r_{11}$, and  $r_{12}$) and Maxwell (negative $r_1$
semi-axis, $r_{13}^+$, and $r_{13}^-$) sets were obtained. They are all
indicated in Fig. \ref{fig-cast1}.
In this
representation it is required that the denominator in (\ref{r1}) and
(\ref{r2}) is positive. The separatrix for $r_1>0$ is associated to
minima while for $r_1<0$ is associated to maxima 
(except the negative $r_1$ semi-axis). In order to clarify
the figure on the separatrix, the energy surfaces corresponding to each
set are plotted as insets. The half plane with $r_2>0$ corresponds to
prolate nuclei, while the one with $r_2<0$ corresponds to oblate
nuclei. Note that 
expressions (\ref{r1p}) and (\ref{r2p}) are only valid for prolate
nuclei, but can be readily obtained for the oblate case.  
On this figure the symmetry limits and
the correspondence with Casten's  
triangle \cite{Iach87} are also represented.
For completeness one should consider the case where the
denominator of (\ref{r1}) and (\ref{r2}) is negative. It implies
that the energy scale becomes negative and the energy surface should
be inverted. The separatrix for this case is plotted in figure
\ref{fig-cast2} and corresponds to the inversion of figure
\ref{fig-cast1}. Again the schematic energy surfaces corresponding to
each branch of the separatrix are shown as insets. Note that in this
case the symmetry limits do not appear in the figure because they
correspond to positive denominators for  $r_1$ and $r_2$. In our
analysis only prolate nuclei are considered, because of that a new figure,
Fig.~\ref{fig-cast-conv}, is included.
In this figure, the right panel corresponds to positive denominators for 
$r_1$ and $r_2$ while the left panel shows the case of negative denominator for
$r_1$ and $r_2$.  In the following we will follow the convention
presented in this figure.

A set of parameters in the Hamiltonian corresponds to a point in the
separatrix plane. The location of the point in that plane provides
the required information on its transitional phase character. As
mentioned above, it follows that  points located on a separatrix line 
correspond to
critical points. Note that  the
dynamical behavior of the system is controlled by the lowest minimum
in the energy surface.  
In this sense we are adopting the Maxwell convention in  
the catastrophe theory language \cite{Gilm81} and the only relevant
branches of the separatrix are $r_{13}^+$ and $r_2=0$ with
$r_1 \le 0$. All these branches correspond to first order phase 
transitions except for 
the single point $(r_1=0, r_2=0)$ that corresponds
to a second order phase transition. The rest of Maxwell lines do not
correspond to a phase transition because they are related to maxima.
The interest of the bifurcation set, corresponding to minima, 
arises  from the fact that it  defines regions where two minima exits. 
In the following subsection the
transitional isotope chains studied in this paper are analyzed in the
separatrix plane.

\subsection{Rare-earth region on the separatrix plane}
\label{sec-represent}
The fits presented in Sect.~\ref{sec-fits} provide  the parameter
sets given in Tables \ref{tab-ham0} and \ref{tab-ham} 
for the four isotope chains studied
in this paper. In this section we plot the corresponding sequences 
of points representing the isotopes in each chain
on the separatrix plane. As can be observed in the previous tables all
the parameters for each chain are fixed except the value of
$\varepsilon_d$ that changes along the chain.

\begin{figure}[tbt]
\begin{center}
\mbox{\epsfig{file=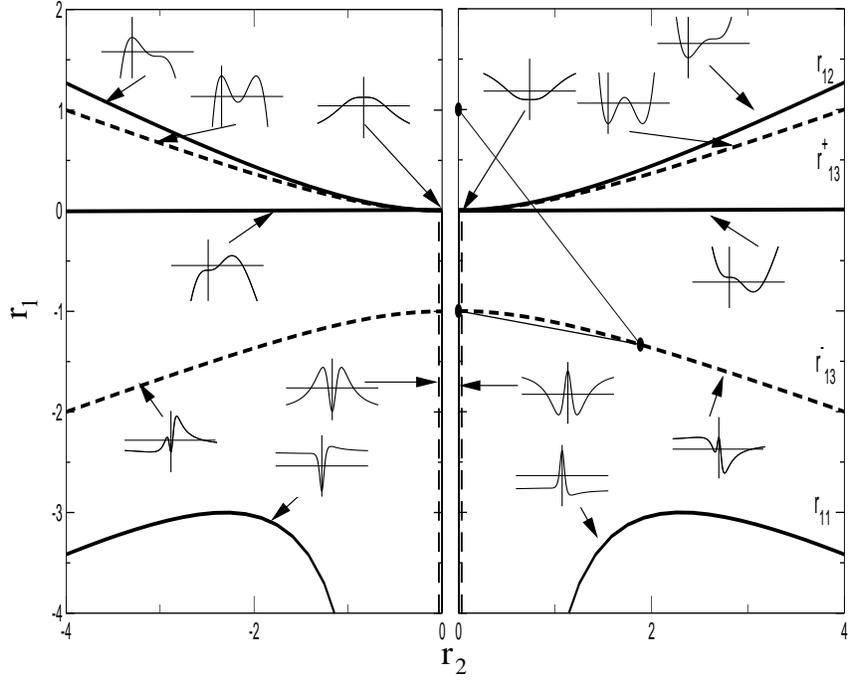,height=9.0cm,angle=0}}
\end{center}
\caption{Separatrix plane for prolate nuclei ($\chi<0$).}
\label{fig-cast-conv}
\end{figure}

\begin{figure}[tbt]
\begin{center}
\mbox{\epsfig{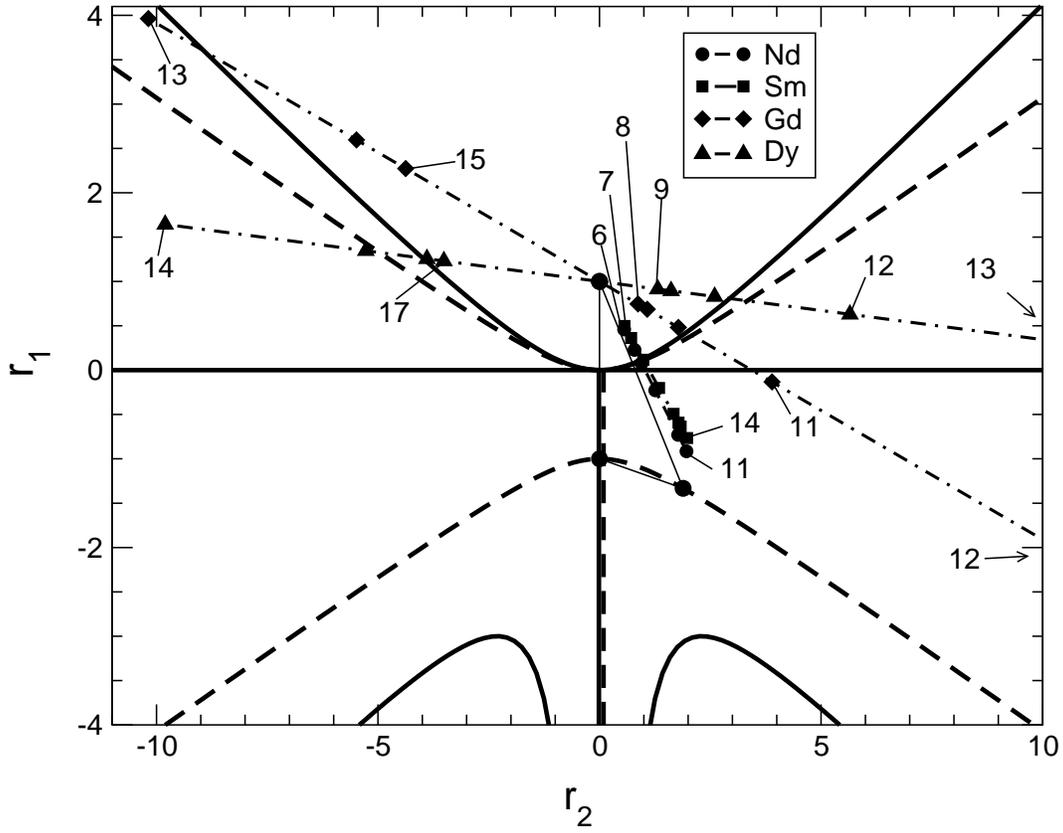}}
\end{center}
\caption{Representation of isotopes in the separatrix plane 
(with $\chi<0$).The numbers on the isotopes correspond to the number
  of bosons.}
\label{fig-iso-all}
\end{figure}

In figure \ref{fig-iso-all} the
positions of the different isotopes in the chains studied are plotted
in the separatrix plane. The interpretation of these  lines is given in  
Fig.~\ref{fig-cast-conv}. As mentioned above, all isotopes in a chain
lie on a straight line. The lighter ones are close to the $U(5)$ point
(spherical shapes) while as the number of neutrons is increased the
corresponding points get increasingly away.  
For the heavier isotopes of Gd, and Dy the denominator of $r_1$ and
$r_2$ becomes negative, which means that  the left panel in
Fig.~\ref{fig-cast-conv} has to be used.

\begin{figure}[tbt]
\begin{center}
\mbox{\epsfig{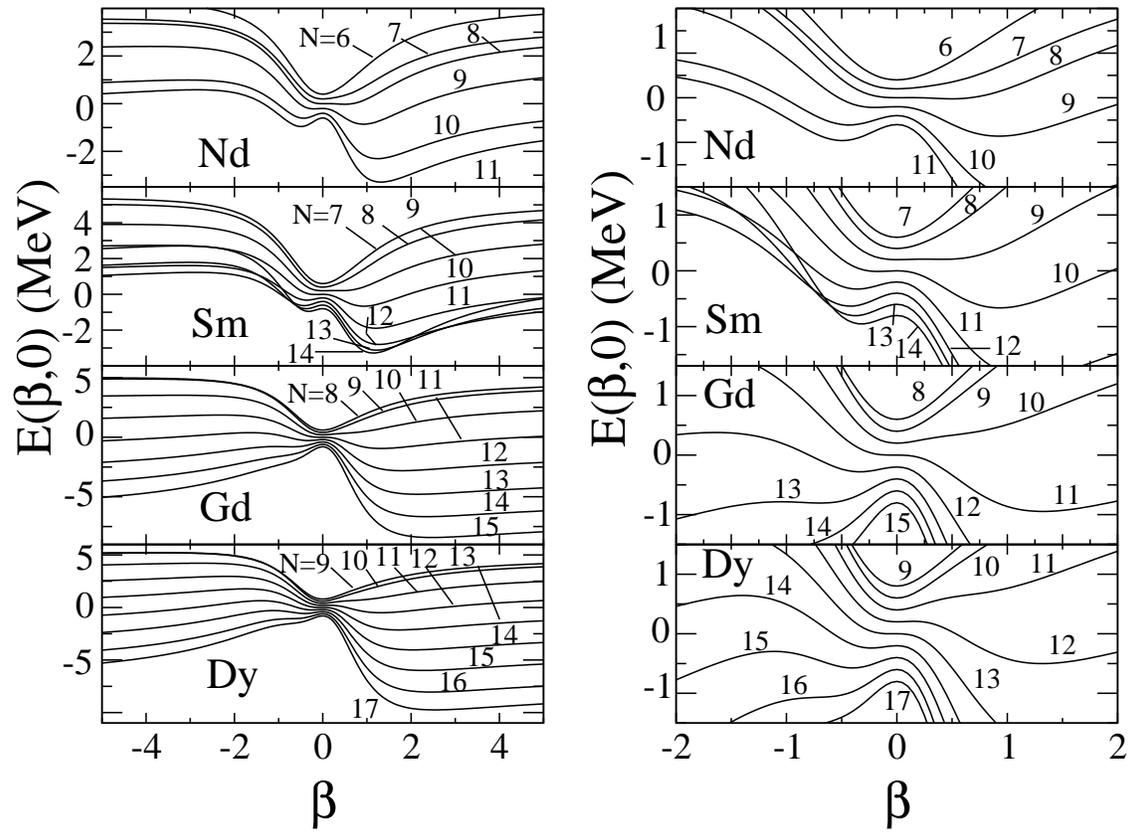}}
\end{center}
\caption{Energy surfaces for the different chain of isotopes.}
\label{fig-ener}
\end{figure}
\begin{figure}[tbt]
\begin{center}
\mbox{\epsfig{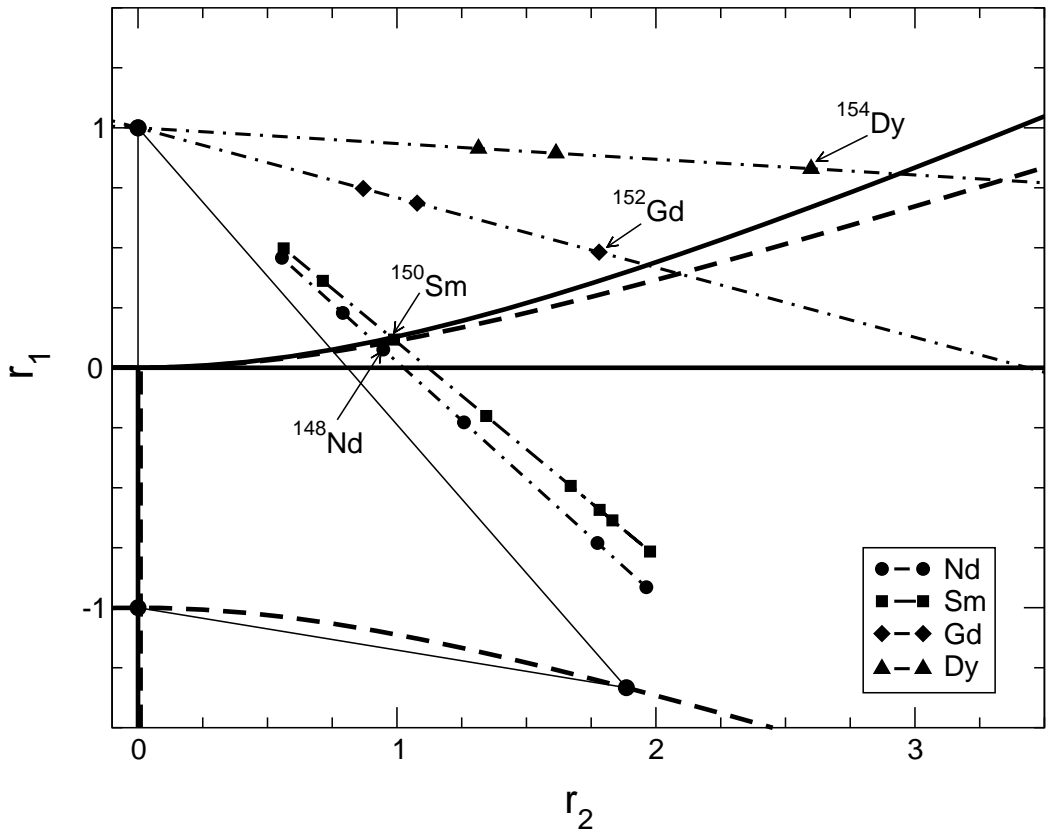}}
\end{center}
\caption{Representation of isotopes in the separatrix plane in a
  closest view.}
\label{fig-iso-all-z}
\end{figure}

The main feature we find is that some nuclei are close to 
the Maxwell set $r_{13}^+$: the closest are $^{148}$Nd (boson number $N=8$)
and $^{150}$Sm (boson number $N=9$) and not far away $^{152}$Gd
(boson number $N=10$). This can be complemented with the image of the
energy surfaces plotted in Fig.  \ref{fig-ener}. The energy surface
for  $^{148}$Nd and $^{150}$Sm are rather flat around $\beta=0$. For  
$^{152}$Gd the situation is not so clear. For Dy there is no isotope
close to the critical point. According to our calculations, the  
transition from spherical to deformed occurs 
 between $N=11$ and $N=12$. The isotope $^{162}$Dy is close to the
Maxwell set but in the left panel. In this situation there should be
two  degenerate maxima. This can be observed in the corresponding 
energy surface (boson number $N=15$) in Fig. \ref{fig-ener}. The
isotopes $^{150}$Nd ($N=9$) and $^{152}$Sm ($N=10$) (also can be
included in this situation $^{154}$Gd ($N=11$) and $^{158}$Dy
($N=13$))  are close to the bifurcation set
$r_2$ axis. Again  inspection of  Fig. \ref{fig-ener} shows that
the energy surfaces for these isotopes has a minimum for $\beta>0$ and
a maximum at  $\beta=0$. In figure \ref{fig-iso-all-z} 
we show an amplification of the critical area. 

In conclusion, from this global  analysis we find that  $^{148}$Nd, $^{150}$Sm,
and (less clearly)  $^{152}$Gd,  are  close to criticality. These
isotopes are quite  close but do not exactly coincide with   
previously proposed
 critical nuclei $^{150}$Nd and $^{152}$Sm 
\cite{Cast01,Kruc02}, where the quite basic criterion was the
closeness of their low-lying  excitation spectra and
transition intensities  with the $X(5)$ values.  

\subsection{Prediction of critical points within CQF}
\label{sec-crit}

The CQF uses a simplified Hamiltonian with only three parameters. For
the description of transitional nuclei from the $U(5)$ to the $SU(3)$
limits the parameters are allowed to vary    nucleus by nucleus. 
The representation of such 
calculations in the separatrix plane shows that all  isotopes in a
chain are basically on top of the straight line connecting the $U(5)$ point,
$(r_1,r_2)=(1,0)$, and the $SU(3)$ point,
$(r_1,r_2)=(-4/3,4\sqrt{2}/3)$. Note that this point corresponds
strictly to the $SU(3)$ Casimir operator. However, a more general CQF
$SU(3)$ Hamiltonian still lies  very close to the latter point. In
general, the same happens in the $U(5)$ and $O(6)$ points.
This means that within this framework the exploration of only a limited 
area in the separatrix plane is allowed. If all  isotopes in an 
isotopic chain are forced to be  located on  the line connecting the 
$U(5)$ and $SU(3)$
points, it follows that  one will more often  find  an isotope close 
to the (unique)  critical point. 
In the calculations presented here we have seen that within the
general formalism this is not always the case. For example, for  Dy 
we did not find   an  isotope close to a  critical point.

In previous systematic studies in the rare-earth region using the 
CQF formalism, Ref.~\cite{Chou97} and \cite{Foss02a}, the
corresponding 
energy surfaces were not presented. We have constructed them
from the parameters given in those references and the
results obtained are consistent with those given in the present
work. In particular, $^{148}$Nd and $^{150}$Sm seem to be  closest
to a  critical point.
 
\section{Conclusions}
\label{sec-conclu}
In this paper we have analyzed  chains of isotopes in the
rare-earth region. In these chains nuclei evolve from spherical to
deformed shapes. We have performed an analysis of the corresponding
shape transitions to look for possible nuclei at or close to a 
critical point. We have used  
the more general one- and two- body IBM Hamiltonian and
generated energy surfaces using the coherent state formalism. 
We have then used  catastrophe theory 
to classify phase transitions and to
decide if a nucleus is close to criticality.  

The approach used to fix the Hamiltonian parameters leads to a very good
global  agreement with the experimental data corresponding to  excitation
energies, B(E2)'s  and S$_{2n}$ values. In particular, an excellent agreement
with the measured S$_{2n}$ values is obtained,  which  is considered  a key
observable to  locate  phase transitional regions. The analysis 
presented here is consistent with previous CQF studies in the same region. As
a result we  find that  $^{148}$Nd and $^{150}$Sm are  the best candidates to
be critical, but we should remark that $^{150}$Nd and $^{152}$Sm are not far
away from it. 

A possible new way
of defining critical nuclei is based on the ``critical symmetries'' $E(5)$ or
$X(5)$ \cite{Iach00,Iach01}. The properties associated with these 
 solutions  allow  the identification of 
critical points by  comparing the experimental data with characteristic 
energy and transition rate ratios.   Thus, 
it may be possible to decide whether a nucleus is critical by
analyzing its spectrum and decay properties. A  trickier  question is 
whether a flat energy surface can be truly associated to a given
nucleus with energy ratios close to $X(5)$. A clear example is
$^{152}$Sm: in section \ref{sec-represent} we have shown that
according to our study the
IBM energy surface of this nucleus is not so flat as expected from 
previous analyses, {\it i.e.} in our work    it does not
correspond to a critical point as  suggested earlier.  However,  if
the spectrum and transition rates are analyzed (see figure
\ref{fig-spect-152sm} and table 
\ref{tab-be2-152sm} ), this nucleus  reproduces reasonably well
the main $X(5)$ features. We note that in the general IBM framework 
there is no unique spectrum associated to a given potential energy
surface,  as implied by equations 
 (\ref{r1}) an (\ref{r2}).  Catastrophe theory constitutes a definite 
criterion regarding this issue, but does not provide a measurable 
signature in itself. 

It seems clear that further  work is required to find experimentally
identifiable features which signal criticality in an unequivocal way.

\section{Acknowledgements}

We acknowledge insightful discussions with R.F.~Casten, F.~Iachello, 
K.~Heyde, J.~Jolie, P.~Van Isacker, and P.~Von Brentano.
This work was supported in part by the Spanish DGICYT under projects
number FPA2000-1592-C03-02 and BFM2002-03315 and by   
CONACYT (M\'exico).


\newpage

\begin{table}
\caption{Values of $\varepsilon_d$ in the Hamiltonian (in keV) for each
  isotopic chain as a function of the neutron number.}
\begin{tabular}{c|ccccccccc}
\hline
 & \multicolumn{9}{c}{Neutron Number} \\
Element & 84 & 86 & 88 & 90 & 92 & 94 & 96 & 98 & 100 \\
\hline
$_{60}$Nd& 1686.3 & 1606.7 & 1645.4 & 1602.9 & 1536.1 & 1595.9 & & &   \\
$_{62}$Sm& 1427.3 & 1393.5 & 1289.3 & 1210.8 & 1158.6 & 1192.5 &
1312.2 & 1452.0 &  \\ 
$_{64}$Gd& 1479.3 & 1508.7 & 1409.0 & 1300.4 & 1221.5 & 1174.4 &
1162.0 & 1176.5 &   \\
$_{66}$Dy& 1558.8 & 1607.6 & 1562.4 & 1503.9 & 1461.0 & 1427.7 &
1413.4 & 1409.2 & 1443.1 \\
\hline
\end{tabular}
\label{tab-ham0}
\end{table}

\begin{table}
\caption{Rest of the parameters in the Hamiltonian  and in the
  E2 transition operator.}
\begin{tabular}{c|cccccccccc|}
\hline
Isotopes & $\tilde{\cal A}$ (MeV) & $\tilde{\cal B}$ (MeV) & $\kappa_0$ (keV)&
$\kappa_1$ (keV) & $\kappa_2$(keV) & 
$\kappa_3$ (keV) &  $\kappa_4$ (keV) & $e_{eff}$ ($e\cdot b$)& $\chi$ \\
\hline
$^{144-154}_{\phantom{4-154}60}$Nd
& 16.75 & -0.51 & 83.753 &
-13.928 & -17.151 & -101.27 & -187.57 & 0.119 & -1.43 \\
$^{146-160}_{\phantom{4-154}62}$Sm
& 18.05 & -0.46 & 53.209 &
-11.267 & -14.674 & -31.769 & -131.24 & 0.119 & -1.69 \\   
$^{148-162}_{\phantom{4-154}64}$Gd
& 22.55 & -0.76 & 45.207 &
-7.932 & -13.129 & -35.224 & -156.24 & 0.110 & -1.77 \\
$^{150-166}_{\phantom{4-154}66}$Dy
& 25.06 & -0.80 & 38.651 &
-6.416 & -13.638 & -59.165 & -163.05 & 0.103 & -1.60 \\
\hline
\end{tabular}
\label{tab-ham}
\end{table}

\begin{table}
\caption{Relevant transition rates for $^{152}$Sm (in w.u.).}
\begin{tabular}{ccccc}
\hline
        & Exp.~ & $X(5)$ & This work & CQF$^{(a)}$   \\
$B(E2:2_1^+\rightarrow 0_1^+)$ &
          144   & 144    & 128       & 144   \\
$B(E2:4_1^+\rightarrow 2_1^+)$ & 
          209   & 228    & 193       & 216   \\ 
$B(E2:6_1^+\rightarrow 4_1^+)$ &
          245   & 285    & 215       & 242   \\
$B(E2:8_1^+\rightarrow 6_1^+)$ &
          285   & 327    & 218       & 248   \\
$B(E2:10_1^+\rightarrow 8_1^+)$ &
          320   & 376    & 210       & 242   \\
$B(E2:0_2^+\rightarrow 2_1^+)$ &
          33    & 91     & 53        & 57    \\    
$B(E2:2_2^+\rightarrow 4_1^+)$ &
          19    & 52     & 14        & 20    \\
$B(E2:2_2^+\rightarrow 2_1^+)$ &
          6     & 13     & 5         & 11    \\
$B(E2:2_2^+\rightarrow 0_1^+)$ &
          1     &  3     & 0         & 0.1   \\
$B(E2:4_2^+\rightarrow 6_1^+)$ &
          4     &  40    & 7         & 14    \\
$B(E2:4_2^+\rightarrow 4_1^+)$ &
          5     &  9     & 2         & 8     \\
$B(E2:4_2^+\rightarrow 2_1^+)$ &
          1     &  13    & 0         & 0.1   \\
\hline
\end{tabular}
\label{tab-be2-152sm}

$^{(a)}$ Following Ref.~\cite{Iach98}.
\end{table}

\newpage

\end{document}